%% file: msapj.tex
\documentclass{aastex}
\usepackage{emulateapj5}

\newcommand{\etal}{{\it et al.}}

\shorttitle{Spectra of Accretion Disks}
\shortauthors{Bhattacharyya et al.}

\begin{document}

\title{General relativistic spectra of accretion disks around rotating neutron
stars}

\author{Sudip Bhattacharyya\altaffilmark{1,2}} \and

\author{Ranjeev Misra\altaffilmark{3}}

\and

\author{Arun V. Thampan\altaffilmark{4}}

\altaffiltext{1}{Joint Astronomy Program, Indian Institute 
of Science, Bangalore 560 012, INDIA; sudip@physics.iisc.ernet.in}
\altaffiltext{2}{Indian Institute of Astrophysics,
Bangalore 560 034, INDIA; sudip@iiap.ernet.in}
\altaffiltext{3}{Department of Physics and Astronomy,
Northwestern University, 2131 Sheridan Road, Evanston,IL 60208;
ranjeev@finesse.astro.nwu.edu}
\altaffiltext{4}{Inter--University Centre for
Astronomy and Astrophysics (IUCAA), Pune 411 007, INDIA;
arun@iucaa.ernet.in}

\begin{abstract}
General relativistic spectra from accretion disks
around rotating neutron stars in the appropriate space-time geometry for several
different equation of state, spin rates and mass of the compact object have been computed.
The analysis involves the computation of the relativistically corrected radial temperature
profiles and the effect of Doppler and gravitational red-shifts on the spectra. 
Light bending effects have been omitted for simplicity. The relativistic spectrum is
compared with the Newtonian one and it is shown that the difference 
between the two is primarily due to the different radial temperature 
profile for the relativistic and Newtonian disk solutions.

To facilitate direct comparison with observations, 
a simple empirical function has been presented which describes the numerically
computed relativistic spectra well. This empirical function (which  
has three parameters including normalization) also describes the 
Newtonian spectrum adequately. Thus the function can in principle be 
used to distinguish between the two. In particular, the best fit 
value of one of the parameters ($\beta$-parameter) $\approx 0.4$ 
for the Newtonian case, while it ranges from $0.1$ to $0.35$ for 
relativistic case depending upon the inclination angle, EOS, spin 
rate and mass of the neutron star. Constraining this parameter by 
fits to future observational data of X-ray binaries will indicate 
the effect of strong gravity in the observed spectrum.
\end{abstract}


\section{Introduction} \label{sec: I} 
X-ray binaries are believed to harbor black holes or weakly magnetized
neutron stars with an accretion disk. The X-ray emission arises from
the hot ($\approx 10^7$ K ) innermost region of the disk. In the
case of a neutron star there will emission, in addition, from a boundary
layer between the accretion disk and neutron star surface. Since the
observed emission arises from regions close to compact object, these sources are
possible candidates for studying strong field gravity. 

In the standard theory (Shakura \& Sunyaev 1973), the accretion disk is
assumed to be an optically thick Newtonian one. 
In this model, the local emergent flux (assumed to be a blackbody) is 
equated to the energy dissipation at a particular radial
point in the disk.  The 
observed spectrum is then a sum of black body components arising from
different radial positions in the disk. General relativistic effects modify 
this Newtonian
spectrum in two separate ways. Firstly, the local energy dissipation at
a radial point is different from the Newtonian disk, giving rise to a modified
temperature profile. Secondly the observed spectrum is no longer a
sum of local spectra because of effects like Doppler broadening, 
gravitational red-shifts 
and light bending. Modified spectra, incorporating these effects, but
with different approximations have been computed by several authors
(e.g. Novikov \& Thorne 1973; Asaoka 1989) for accretion disks around
rotating (Kerr) black holes. These computations confirm the expected
result, that the relativistic spectral shape differs from the Newtonian one
by around 10\%. Thus, for comparison with observed data with systematic and
statistical errors larger than 10\%, the Newtonian
approximation is adequate. Ebisawa, Mitsuda and Hanawa (1991) showed
that for typical data from the Ginga satellite, the relativistic spectrum cannot
be differentiated from the Newtonian disk spectrum. They also found
that the relativistic spectrum is similar in shape (at the sensitivity
level of Ginga ) to the Comptonized model spectrum. Although, Ginga was
not sensitive enough to distinguish between the different spectra, better
estimates of fit parameters like accretion rate and mass of the compact object
were obtained when the data was compared to relativistic spectra rather
than the standard Newtonian one. 

The present and next generation of satellites (e.g. ASCA, RXTE,
Chandra, XMM, Constellation-X)
with their higher sensitivity and/or larger effective area than Ginga are 
expected to differentiate between relativistic and Newtonian spectra from low
mass X-ray binaries (LMXB) and black hole systems. However, as pointed out
by Ebisawa, Mitsuda and Hanawa (1991), the presence of additional components
(e.g. boundary layer emission from the neutron star surface) and smearing 
effects due to Comptonization may make the detection ambiguous. 
Nevertheless, the detection of strong gravity effects on the spectra from
these sources, will be limited by the accuracy of theoretical modeling of
accretion disk spectra rather than limitations on the quality of the
observed data. Thus, it is timely to develop accurate relativistically
corrected spectra for comparison with present and future observations. 
Apart from
the importance of detecting strong gravity effects in the spectra of these
sources, such an analysis may also shed light on the geometry and dynamics
of innermost regions of accretion disks. 

 Novikov \& Thorne (1973) and Page \& Thorne (1974) 
computed the spectra of accretion disks around rotating (Kerr) black
holes.  This formalism when directly applied to rotating neutron stars 
provide only a first order estimate: the absence of an internal solution 
in the case of Kerr geometry, makes it difficult to obtain, in a 
straightforward manner, the coupling between the mass and the angular 
momentum of the central accretor.
This coupling depends on one hand on the equation of state of neutron star
matter and on the other on the proper treatment of rotation within
general relativity.
Equilibrium configurations of rapidly rotating neutron stars for
realistic equations of state have been computed by several authors
(Bonazzola \& Schneider 1974; Friedman, Ipser \& Parker 1986; 
Cook, Shapiro \& Teukolsky 1994; Stergioulas \& Friedman 1994; 
Salgado \etal 1994a,b; Datta, Thampan \& Bombaci 1998).
One crucial feature in all these calculations is that the 
space--time geometry is obtained by numerically and self--consistently 
solving the Einstein equations and the equations for hydrodynamic 
equilibrium for a general axisymmetric metric.  
With the aim of modeling spectra of LMXBs, we attempt, in this paper, to 
compute the spectrum of accretion disks around rotating neutron 
stars within such a space--time geometry. This is particularly important
since LMXBs are old (population I) systems and the central
accretor in these systems are expected to have large rotation rates 
(Bhattacharya \& van den Heuvel 1993 and references therein)

Computation of the spectra is numerically time consuming and hence
direct fitting to the observed data is impractical. For the sake of 
ease in modeling, we also present in this paper, a simple empirical 
analytical expression that describes
the numerically computed spectra. As shown later, the same expression
(which has three parameters including normalization) can also
describe the Newtonian spectra. In particular, the value of one
of the parameters (called $\beta$ parameter here) determines whether the
spectrum is relativistically corrected or not. This will facilitate comparison
with observational data since only this $\beta$-parameter has to be 
constrained to indicate
the effect of strong gravity in the observed spectrum.

The next section describes the method used to compute the spectra. In \S 3
the results of the computation and the empirical fits are shown. \S 4 is
devoted to discussion and summary.

\section{Spectral Computation} \label{sec: 2} 

The disk spectrum is expressed as:
\begin{eqnarray}
F(E_{\rm ob}) & = & (1/E_{\rm ob})\int I_{\rm ob}(E_{\rm ob}) d\Omega_{\rm ob}
\label{eq:feorb}
\end{eqnarray}
\noindent where the subscript `ob' denotes the quantity in observer's frame, 
the flux $F$ is expressed in photons/sec/cm$^2$/keV, $E$ is photon energy 
in keV, $I$ 
is specific intensity and $\Omega$ is the solid angle subtended by the source 
at the observer. 

As $I/E^3$ remains unchanged along the path of a photon (see for e.g.,
Misner et al. 1973), 
one can calculate $I_{\rm ob}$, if $I_{\rm em}$ is known (hereafter, the 
subscript `em' denotes the quantity in emitter's frame). We assume the
disk to emit like a diluted blackbody, so
$I_{\rm em}$ is given by 
\begin{eqnarray}
I_{\rm em} & = & (1/f^4) B(E_{\rm em},T_{\rm c})
\end{eqnarray}
\noindent where $f$ is the color factor of the disk assumed to
be independent of radius (e.g. Shimura \& Takahara 1995). $B$ is the 
Planck function and $T_{\rm c}$ (the temperature in the central plane of
the disk), is related to the effective temperature $T_{\rm eff}$ through
the relation $T_{\rm c}= f T_{\rm eff}$. The effective temperature,
$T_{\rm eff}$ is a function of the radial coordinate $r$ 
and for a rotating accretor, is given by (Page \& Thorne 1974) 
\begin{eqnarray}
T_{\rm eff} & = & (F/\sigma)^{1/4}
\end{eqnarray}
\noindent where 
\begin{eqnarray}
F(r) & = & -\frac{\dot{M}}{4 \pi r} \Omega_{{\rm K},r} 
(\tilde{E} - \Omega_{\rm K} \tilde{l})^{-2}
\int_{r_{\rm in}}^{r} (\tilde{E} - \Omega_{\rm K} \tilde{l}) \tilde{l}_{,r} dr
\end{eqnarray}
\noindent is the flux of energy from the disk in orbiting 
particle's frame. The factor $\sigma$ is the Stephan--Boltzmann constant, 
$r_{\rm in}$ is the disk inner edge radius, $\tilde{E}$, $\tilde{l}$
are the specific energy and specific angular momentum of a test particle
in a Keplerian orbit and $\Omega_{\rm K}$ is the Keplerian angular velocity at
radial distance $r$. In our notation, a comma followed by a variable as
subscript to a quantity, represents
a derivative of the quantity with respect to the variable. Also, in this
paper, we use the geometric units $c=G=1$.

The quantities $E_{\rm ob}$ and $E_{\rm em}$ are related through the
expression $E_{\rm em}$ = $E_{\rm ob} (1 + z)$, where $(1 + z)$ contains the 
effects of both gravitational redshift and Doppler shift. For a
general axisymmetric metric (representing the space--time geometry
around a rotating neutron star), the factor $(1 + z)$ is expressed as 
(see for e.g. Luminet 1979) 
\begin{eqnarray}
1 + z & = & (1 + \Omega_{\rm K} b \sin{\alpha} \sin{i}) {(-g_{tt} - 2 
\Omega_{\rm K} g_{t\phi} - {\Omega_{\rm K}}^2 g_{\phi\phi})}^{-1/2}
\label{eq:1pz}
\end{eqnarray}
\noindent where the $g_{\mu\nu}$'s are the metric coefficients and $t$
and $\phi$ are the time and azimuthal coordinates. In the above 
expression (which includes light--bending effects), $i$ is the
inclination angle of the source, $b$ the impact
parameter of the photon relative to the line joining the source 
and the observer and $\alpha$ the polar angle of the position of the
photon on the observer's photographic plate.
For the sake of illustration and simplicity in calculations, we
neglect light bending.  We thus write 
$b \sin{\alpha}$ = $r \sin{\phi}$ and
\begin{eqnarray}
d\Omega_{\rm ob} & = & r dr d\phi \cos{i} \over D^2
\end{eqnarray}
\noindent where $D$ is the distance of the source from the observer.

The space--time geometry around a rotating neutron star can be
described by a general axisymmetric, stationary metric (see for e.g. Komatsu, 
Eriguchi \& Hachisu 1989).  Assuming the matter to be a perfect
fluid, and the metric to be asymptotically flat, the Einstein field
equations reduce to three non--homogeneous, second--order, coupled
differential equations and one ordinary differential equation in
terms of the energy density and the pressure of the high--density
neutron star matter (Cook, Shapiro \& Teukolsky 1994).  An important
input in solving these equations is the equation of state (EOS) 
of high density matter composing the neutron star.  Assuming
rigid rotation, we solve these equations numerically and
self--consistently for four representative EOS models:
(A) Pandharipande (1979) (hyperons), (B) Baldo, Bombaci \& Burgio (1997) 
(AV14 + 3bf), (C) Walecka (1974) and (D) Sahu, Basu \& Datta (1993). The
values of stiffness parameters for each of these EOS models are widely
different and increases from (A) to (D).  We, therefore, expect the
results of our computations to be of sufficient generality.

The solution of the field equations yield the metric coefficients,
numerically, as functions of $r$ and $\theta$.  Using these metric
coefficients, it is straightforward to calculate the structure
parameters of rapidly rotating neutron stars.  The details of these 
calculations are given in Datta, Thampan \& Bombaci (1998) and references 
therein.  The quantities: $r_{\rm in}$, $\tilde{E}$, $\tilde{l}$ and
$\Omega_{\rm K}$ are obtained by solving the equation of motion of 
material particles within the space--time geometry given by the
above metric (Thampan \& Datta 1998; Bhattacharyya \etal 2000).
For our purpose here, we compute constant gravitational mass
sequences whose rotation rates vary from zero to the centrifugal mass
shed limit (where gravitational forces balance centrifugal forces). 
For realistic neutron stars, the inner radius $r_{\rm in}$ may be located
either at the marginally stable orbit or the surface of the neutron
star depending on its central density and rotation rate 
(Thampan \& Datta 1998; Bhattacharyya \etal 2000), having important 
implications for the gravitational energy release as well as the 
temperature profiles of accretion disks.
For the procedure of calculating $T_{\rm eff}$, for rapidly rotating neutron 
stars, considering the full effect of general relativity, we refer to 
Bhattacharyya et al., 2000. These authors have also shown (in their Fig. 2) 
that the difference between Newtonian temperature profile and general 
relativistic temperature profile is substantial at the inner portion of the 
disk.  As will be shown herein, it turns out that this is the major reason 
for the difference between Newtonian and general relativistic spectra at 
high energies.

To summarize this section, we calculate the accretion disk spectrum 
using equation~\ref{eq:feorb}, taking the radial integration limits as 
$r_{\rm in}$ 
and $r_{\rm out}$ and the azimuthal integration limits as 0 and 2$\pi$. 
We choose a very large value ($\approx 10^5$ Schwarzschild radius) for 
$r_{\rm out}$. 

\section{Results} \label{sec: 3} 

To illustrate the differences between the relativistic and 
Newtonian spectra, we show in Figure~1, the computed relativistic 
spectrum (solid line) and the Newtonian spectrum (dotted line) for 
the same parameters. The Newtonian spectrum is the spectrum
expected from a standard non-relativistic disk (Shakura
\& Sunyaev 1973) but with the intensity and the effective temperature
modified by the color factor (equations 2 and 3).
In order to isolate the different contributions, 
we have also plotted in Figure~1, the theoretical spectrum arising 
from relativistic temperature profile but without the effect of 
Doppler/gravitational red-shift (dashed line).  The relativistic 
spectrum is under-luminous compared to the Newtonian one at high 
energies -- this is primarily because of the the difference in the
radial temperature profile (Bhattacharyya \etal 2000). The difference 
between the two spectra is nearly 50\% at 2 keV.  We emphasize here, 
that such high difference is only true when both the spectra are 
calculated for the same disk parameters. If the Newtonian spectra 
is calculated for slightly different values of disk parameters (e.g. 
accretion rate, inclination angle, distance to the source) the 
average discrepancy between the two spectra will be less (Ebisawa, 
Mitsuda and Hanawa 1991).

In order to facilitate comparison with observations, we introduce a simple
analytical expression which empirically describes the computed relativistic
(and Newtonian) spectra.
\begin{eqnarray}
S_{\rm f} ( E ) & = & S_o E_a^{-2/3} ({E\over E_a})^{\gamma} exp(-{E\over E_a})  
\end{eqnarray}
where, $\gamma = -(2/3)(1+E\beta/E_a)$, $E_a$, $\beta$ and $S_o$ are 
parameters and $E$ is the energy of the photons in keV. $S_{\rm f} 
(E) $ is in units of photons/sec/cm$^2$/keV. To compare this empirical 
function with the computed spectra, we use a reduced $\chi^2$ technique. 
In particular, we define a function
\begin{eqnarray}
\chi^2 & = & {1\over N}\sum_{i = 1}^{N} ({S_{\rm c}(E_i)-S_{\rm f}(E_i) 
\over 0.1  S_{\rm c}(E_i)})^2   
\end{eqnarray}
where $S_{\rm c}(E)$ is the computed spectra. The spectra are divided 
into $N$ logarithmic energy bins. We have chosen the range of energy 
used in calculating $\chi^2$ to be dependent on the location of the 
maximum of the energy spectrum ( $E S_{\rm c} (E)$) which is typically 
at 2 keV. The minimum energy is set to be  one hundredth of this value 
(typically 0.02 keV) while the maximum is set at ten times (typically 
20 keV). $\chi^2$ is fairly insensitive to the number of energy bins; 
we take $N = 200$. For each $S_{\rm c}(E)$ the best fit parameters 
($E_a$, $\beta$ and $S_o$) are obtained by minimizing $\chi^2$.  
Figure~2 shows the relativistic spectra for three different inclination 
angles (solid lines) and the corresponding empirical fits using 
equation~(\ref{eq:feorb}) (dotted lines). The minimum $\chi^2$ obtained 
while fitting these spectra was $ < 0.1$, which means that the average
discrepancy is less than 3 \%. This is also true for other disk parameters
and EOS considered in this work. Thus the empirical function 
(eq.~[\ref{eq:feorb}]) is a reasonable approximation to the computed 
relativistic spectra. It also describes the Newtonian spectra to a 
similar degree of accuracy.

The $S_o$ parameter in equation~(\ref{eq:feorb}), is the normalization 
factor and is independent of the relativistic effects. It depends only 
on the mass of the star ($M$), accretion rate ($\dot M$), distance to 
the source ($D$), color factor (f) and inclination angle (i) i.e 
$S_o \propto \dot M^{2/3} f^{-4/3} M^{1/2} D^{-2} \cos{i}$.  The $E_a$ 
parameter (which is in units of keV) describes the high energy cutoff 
of the spectrum. Its dependence on the space time metric and inclination 
angle is complicated but it scales as $E_a \propto \dot M^{1/4} f$. The 
$\beta$-parameter depends only on the space time metric and not on 
accretion rate, distance to the source or color factor. This makes
the $\beta$-parameter useful as a probe into the underlying space-time 
metric.  We show in Figure~3, the variation of minimum $\chi^2$ 
(i.e. minimized w.r.t. to parameters $E_a$ and $S_o$ only) as a 
function of the $\beta$-parameter for the three spectra shown in 
Figure~2 and for the Newtonian one. For the Newtonian case the minimum 
$\chi^2$ occurs for $\beta \approx 0.4$ while it is lower
for the relativistic cases. For example, consider the relativistic
spectrum for parameters listed in Figure~1 and for $i = 30^o$ (line
marked as 3 in Figure~3). If this spectrum is fitted with the empirical
function the minimum $\chi^2 = 0.05$ (corresponding to an average 
discrepancy of 2\%) and the best fit $\beta$-parameter is 
$\beta \approx 0.25$. For a Newtonian $\beta$-parameter value of 
$\approx 0.4$, the minimum $\chi^2$ increases to  0.1, corresponding 
to an average discrepancy of more than 3\%. Thus the empirical function can 
resolve the difference between the Newtonian and the relativistic 
one at the 10\% level.  For an observed spectrum fitted using the 
empirical function, if the best fit range of $\beta$-parameter 
excludes the Newtonian value of 0.4, that would strongly indicate 
that the spectrum has been modified by strong gravitational effects.  
Since the $\beta$-parameter deviation from 0.4 increases with 
inclination angle, nearly edge on disks are more promising 
candidates for detecting the presence of strong gravity. To show 
the robustness of this result we show in Figure~4, 5 and 6 the 
variation of the best fit $\beta$-parameter with $i$ for different 
EOS, mass of the central object and spin rates respectively.  For 
all these cases the best fit $\beta$-parameter is less than 0.4.  
Parameter $E_a$ is useful to determine the accretion rate. However, 
it also depends on the metric and inclination angle. We show this 
dependence in Figure~7.

\section{Summary and discussion} \label{sec: 4} 

In this paper, we have computed relativistic spectra from 
accretion disks around rotating neutron stars for the appropriate 
space-time geometry. Several different EOS, spin rates and mass of 
the compact object have been considered.  The Doppler and 
gravitational effects on the spectra have been taken into account,
while light bending effects have been omitted for simplicity. The 
spectrum differs from the Newtonian one, with the main difference 
being due to the different radial temperature profile for the 
relativistic and Newtonian disk solutions. 

A simple empirical function has been presented which describes 
the numerically computed relativistic spectra well. This will 
facilitate direct comparison with observations. The empirical 
function (eq.~[7]) has three parameters including normalization. 
Another important advantage of this function is that it also 
describes the Newtonian spectrum adequately, and the value of one
of the parameters ($\beta$-parameter) distinguishes between the two. 
In particular, the best fit $\beta$-parameter $\approx 0.4$ for the 
Newtonian case, while it ranges from $0.1$ to $0.35$ for relativistic 
case depending upon the inclination angle, EOS, spin rate and mass 
of the neutron stars. 

In principle, for sufficiently high quality data, the effects of
strong gravity on the disk spectrum can be detected using this empirical
function as a fitting routine and constraining the $b$-parameter. 
However, it must be emphasized that there are several reasons 
why this may not be possible. There could be systems which have 
additional components in the X-ray spectra; for example boundary 
layer emission from the neutron star surface. Uncertainties
in modeling these additional components may lead to a wider range 
in the best fit $\beta$-parameter. Thus accurate spectra of the 
boundary layer (with relativistic corrections) is also needed for 
modeling these systems . Moreover, X-rays could be emitted from hotter 
regions (e.g an innermost hot disk or a corona) giving rise to a 
Comptonized spectra instead of the sum of local emission assumed 
here. In this case, the empirical fit will probably not describe 
the observed data well. It has been assumed here that
the color factor is independent of radius. 
Shimura and Takahara (1995) have shown from numerical computation that this
could be the case for an accretion disk in a
Schwarzschild metric. Apart from the fact that
this was done for Schwarzschild metric, their numerical
computation depends on the vertical structure of the
disk which in turn depends on the unknown viscosity
mechanism in the disk. If the color
factor has a radial dependence, 
the spectral shape might change which may be confused to be a 
relativistic effect. 

Despite these caveats the method described in this paper will be 
a step forward in the detection of strong gravity effects in the 
spectra of X-ray binaries. Future comparison with high quality 
observational data, will highlight the theoretical requirements 
that have to be met, before concrete evidence for strong gravity  
are detected in these systems and the enigmatic region around 
compact objects is probed.  

\acknowledgements

The authors acknowledge suggestions from the late Professor Bhaskar
Datta in early stages of this work.  They also thank
Dipankar Bhattacharya for suggesting improvements in the
presentation of the manuscript. SB thanks Pijush Bhattacharjee 
for encouragement.

\newpage
\input psbox.tex

\newpage
\begin{figure*}[h]
\hspace{-1.5cm} 
{\mbox{\psboxto(17cm;20cm){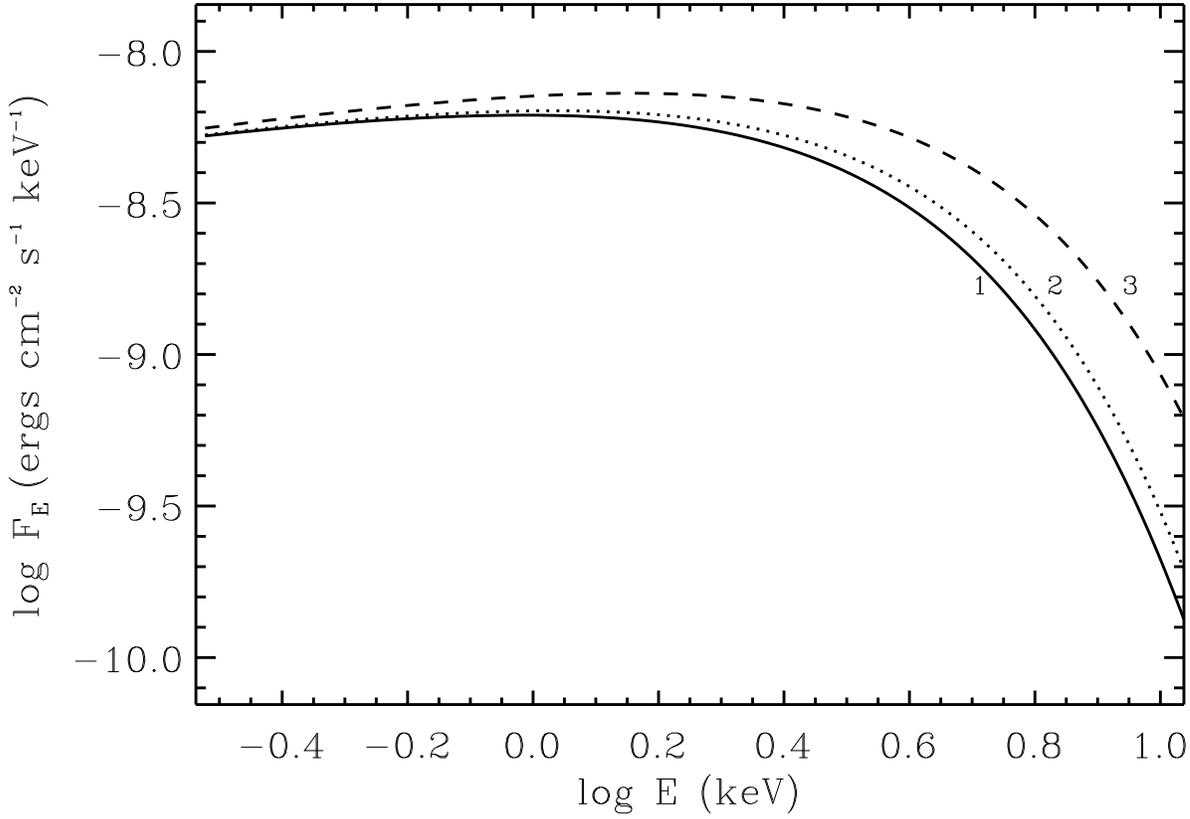}}}
\caption{\label{fig:1}General relativistic spectrum (solid line) for a 
neutron star 
configuration with mass $M = 1.4 M_\odot$, spin rate $\Omega_* = 0$, 
distance to the source $D = 5$ kpc, inclination angle $i = 30^o$, 
accretion rate $\dot M = 10^{18}$ g/sec and color factor $f = 2$. Dashed 
line: the spectrum expected from a source with the same disk parameters
but without the relativistic effects (Newtonian spectrum). Dotted 
line: The spectrum for the same disk parameters but without the effect 
of Doppler and gravitational red-shifts (i.e. $z$ is set to zero). 
The EOS model (B) is used here.}
\end{figure*}
\newpage
\begin{figure*}[h]
\hspace{-1.5cm}
{\mbox{\psboxto(17cm;20cm){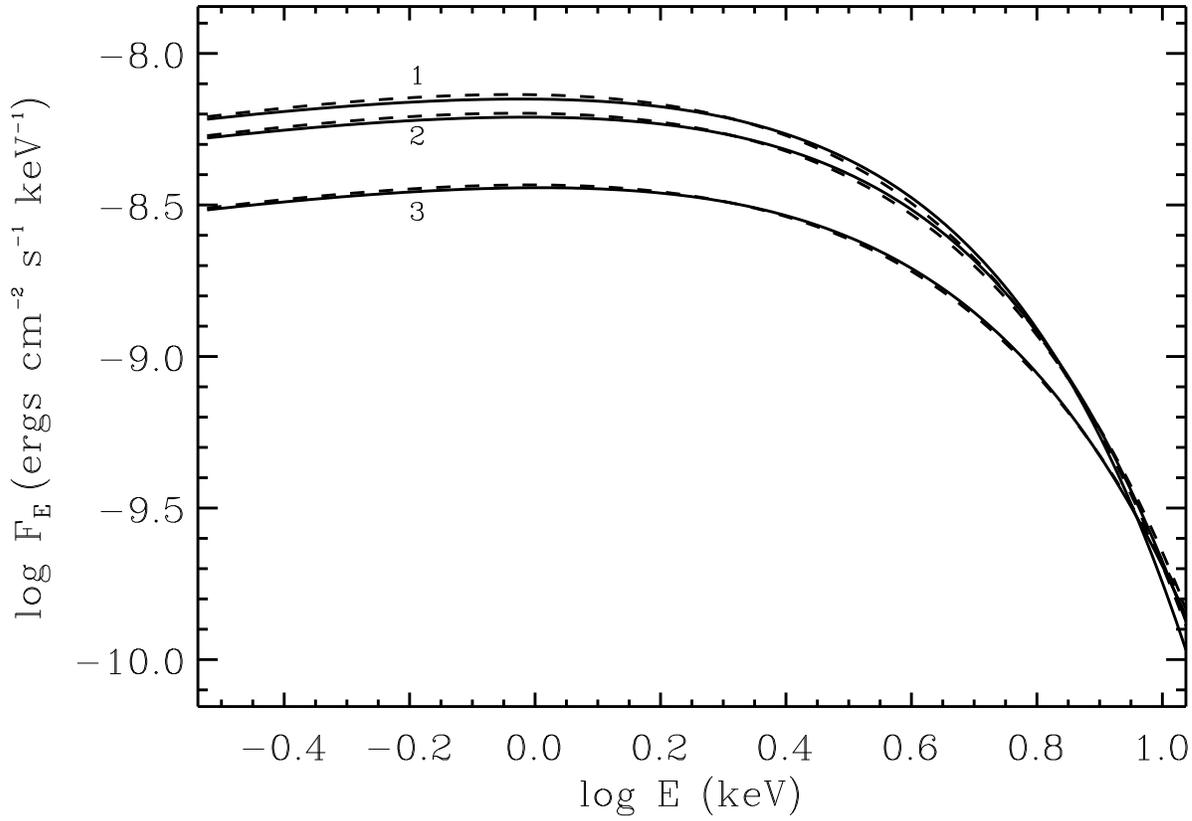}}}
\caption{\label{fig:2}The relativistic spectra for three different 
inclination angles ( $ i = 0^o,30^o,60^o$ ) with rest of the parameters same 
as in Figure~\ref{fig:1} (solid lines). Dashed lines: empirical fit to
the relativistic spectra using equation~(7). The minimum 
$\chi^2 = 0.073,0.049,0.026$ for $i = 0^o,30^o,60^o$ respectively. }
\end{figure*}

\newpage
\begin{figure*}[h]
\hspace{-1.5cm}
{\mbox{\psboxto(17cm;20cm){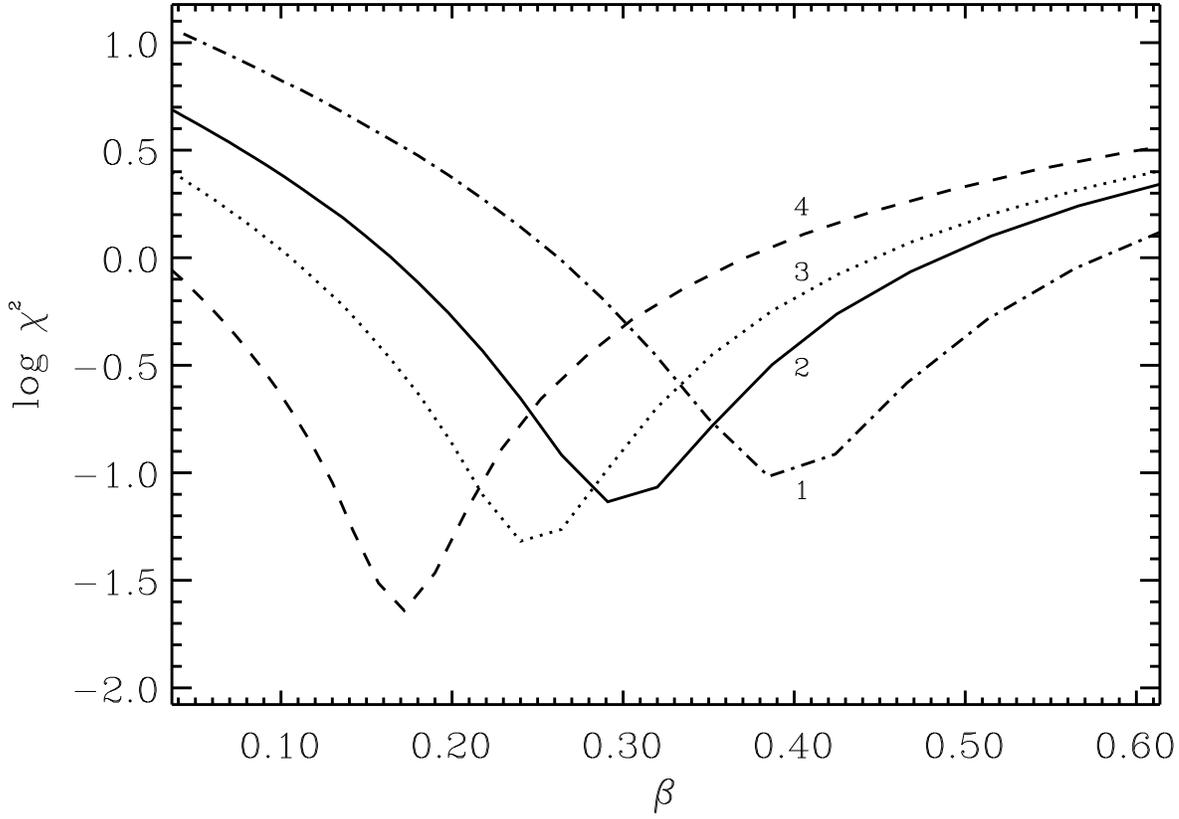}}}
\caption{\label{fig:3}Variation of minimum $\chi^2$ (i.e. minimized w.r.t to 
parameters $E_a$ and $S_o$) with parameter $\beta$. Curves marked 2, 3 
and 4 correspond to  the spectra shown in Figure~\ref{fig:3} 
for $i = 0^o,30^o,60^o$ respectively. Curve marked 1 is for the 
Newtonian spectra shown in Figure~\ref{fig:1}.}
\end{figure*}

\newpage
\begin{figure*}[h]
\hspace{-1.5cm}
{\mbox{\psboxto(17cm;20cm){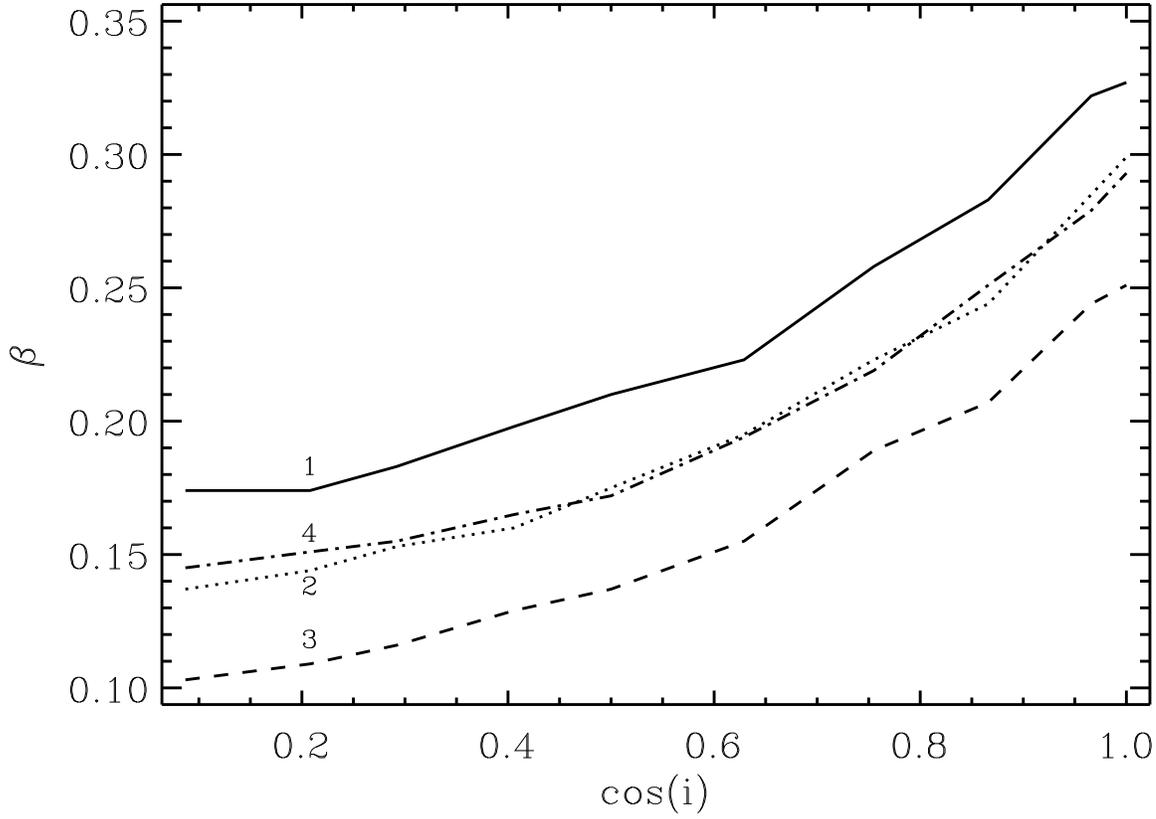}}}
\caption{\label{fig:4}Variation of the best-fit $\beta$-parameter with 
inclination 
angle for different equation of states. Curve 1:Pandharipande(Y) 
(softest), curve 2: Bombaci, curve 3: Walecka curve 4: SBD (stiffest).
The values of the other parameters are as in Figure~\ref{fig:1}.}
\end{figure*}

\newpage
\begin{figure*}[h]
\hspace{-1.5cm}
{\mbox{\psboxto(17cm;20cm){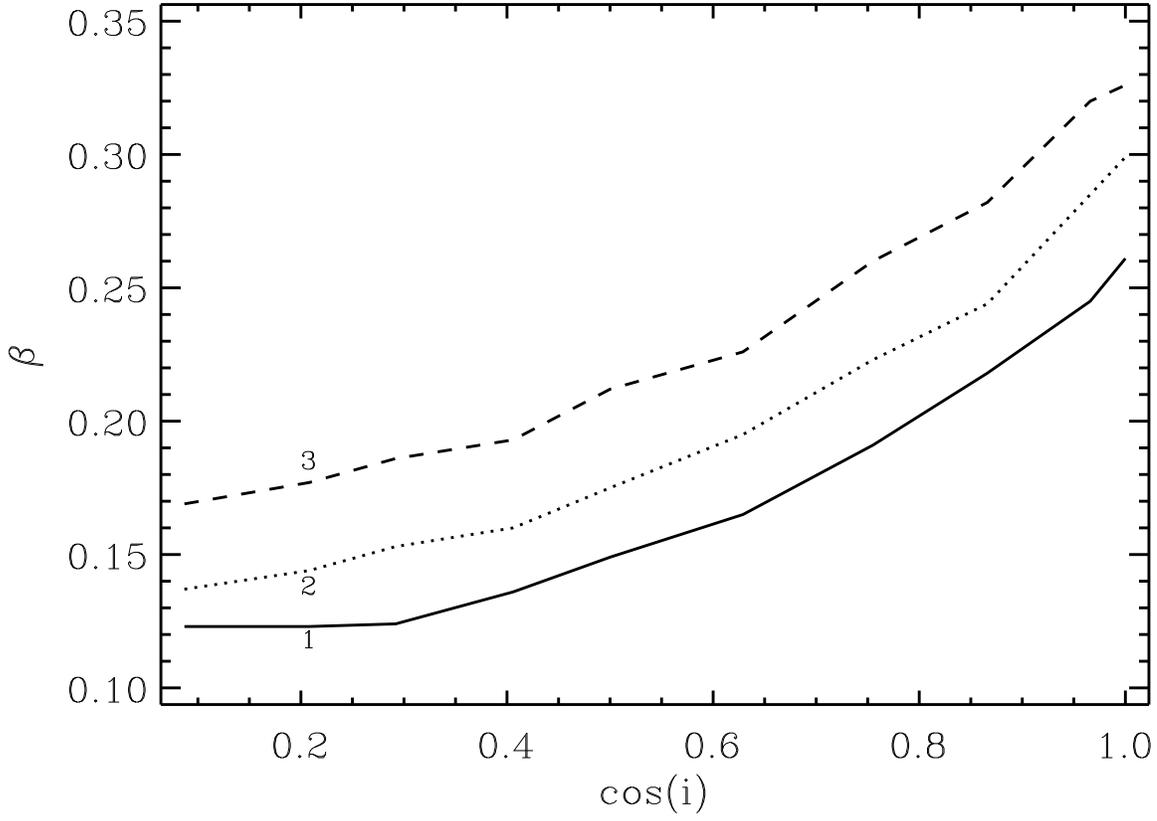}}}
\caption{\label{fig:5}Variation of the best-fit $\beta$-parameter with 
inclination angle 
for different neutron star masses. Curve 1: $ M = 1.0 M_\odot$, 
curve 2: $ M = 1.4 M_\odot$,
curve 3: $ M = 1.788 M_\odot$.
The values of the other parameters are as in Figure~\ref{fig:1}.} 
\end{figure*}

\newpage
\begin{figure*}[h]
\hspace{-1.5cm}
{\mbox{\psboxto(17cm;20cm){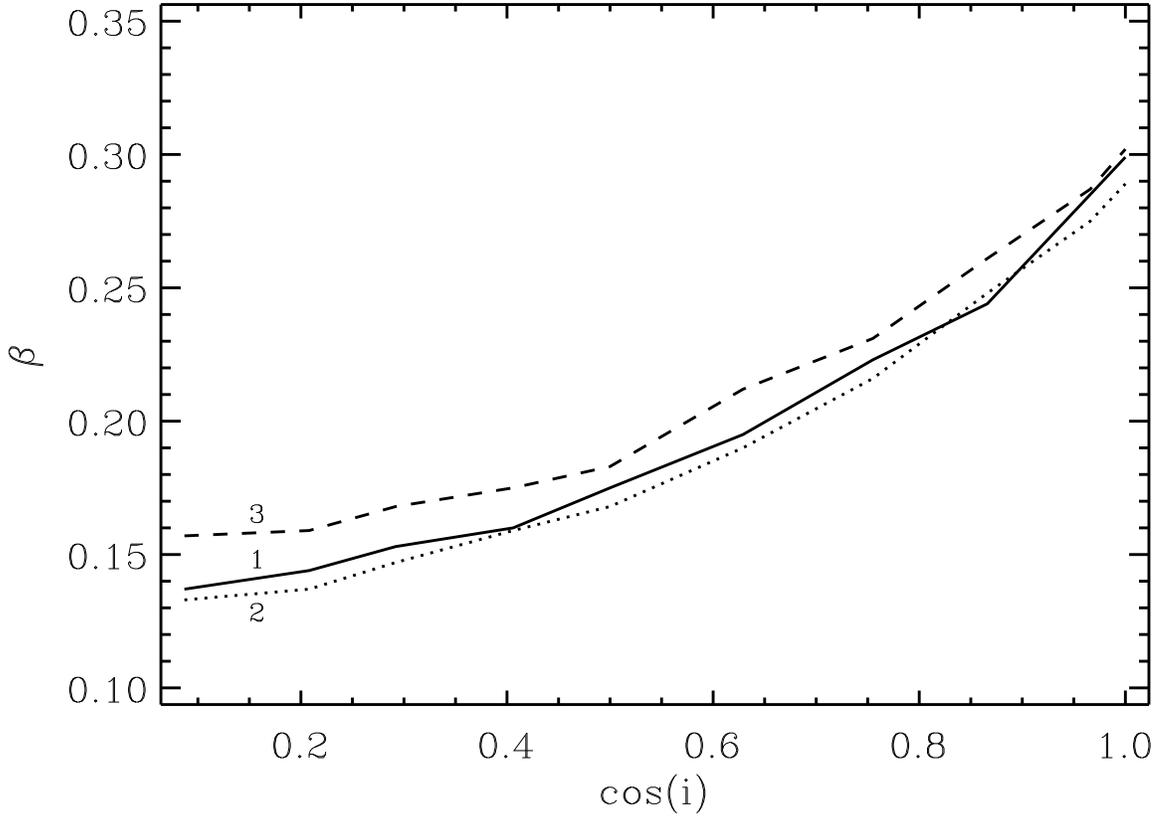}}}
\caption{\label{fig:6}Variation of the best-fit $\beta$-parameter with 
inclination 
angle for different neutron star spin rates. Curve 1: 
$ \Omega_* = 0$ radians/sec, curve 2: $ \Omega_* = 2044$ 
radians/sec, curve 3: $ \Omega_* = 7001$ radians/sec 
(mass-shed limit).
The values of the other parameters are as in Figure~\ref{fig:1}.}
\end{figure*}

\newpage
\begin{figure*}[h]
\hspace{-1.5cm}
{\mbox{\psboxto(17cm;20cm){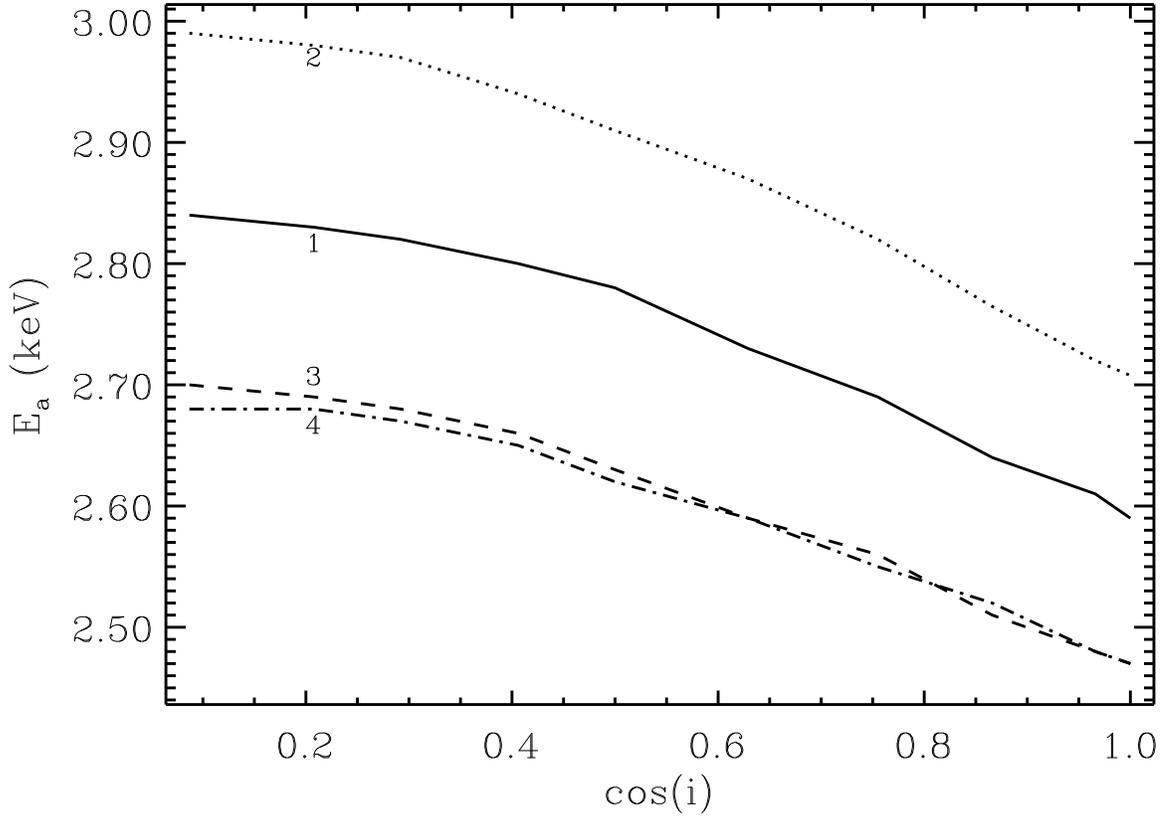}}}
\caption{\label{fig:7}Variation of the best-fit $E_a$-parameter with 
inclination 
angle for different equation of states. The curve numbers correspond 
to the same equation of states as listed in Figure~\ref{fig:4}.
The values of the other parameters are as in Figure~\ref{fig:1}.} 
\end{figure*}
\end{document}

%% file: psbox.tex
%
%
%
%
%
\def\temp{1.34}%
\let\tempp=\relax
\expandafter\ifx\csname psboxversion\endcsname\relax
  \message{PSBOX(\temp) loading}%
\else
    \ifdim\temp cm>\psboxversion cm
      \message{PSBOX(\temp) loading}%
    \else
      \message{PSBOX(\psboxversion) is already loaded: I won't load
        PSBOX(\temp)!}%
      \let\temp=\psboxversion
      \let\tempp= 
    \fi
\fi
\tempp
\let\psboxversion=\temp
\catcode`\@=11
%
%
\def\psfortextures{
\def\PSspeci@l##1##2{%
\special{illustration ##1\space scaled ##2}%
}}%
\def\psfordvitops{
\def\PSspeci@l##1##2{%
\special{dvitops: import ##1\space \the\drawingwd \the\drawinght}%
}}%
\def\psfordvips{
\def\PSspeci@l##1##2{%
\d@my=0.1bp \d@mx=\drawingwd \divide\d@mx by\d@my
\includegraphics{##1\space}}}%
\def\psforoztex{
\def\PSspeci@l##1##2{%
\special{##1 \space
      ##2 1000 div dup scale
      \number-\psllx\space \number-\pslly\space translate
}}}%
\def\psfordvitps{
\def\psdimt@n@sp##1{\d@mx=##1\relax\edef\psn@sp{\number\d@mx}}
\def\PSspeci@l##1##2{%
\special{dvitps: Include0 "psfig.psr"}
\psdimt@n@sp{\drawingwd}
\special{dvitps: Literal "\psn@sp\space"}
\psdimt@n@sp{\drawinght}
\special{dvitps: Literal "\psn@sp\space"}
\psdimt@n@sp{\psllx bp}
\special{dvitps: Literal "\psn@sp\space"}
\psdimt@n@sp{\pslly bp}
\special{dvitps: Literal "\psn@sp\space"}
\psdimt@n@sp{\psurx bp}
\special{dvitps: Literal "\psn@sp\space"}
\psdimt@n@sp{\psury bp}
\special{dvitps: Literal "\psn@sp\space startTexFig\space"}
\special{dvitps: Include1 "##1"}
\special{dvitps: Literal "endTexFig\space"}
}}%
\def\psfordvialw{
\def\PSspeci@l##1##2{
\special{language "PostScript",
position = "bottom left",
literal "  \psllx\space \pslly\space translate
  ##2 1000 div dup scale
  -\psllx\space -\pslly\space translate",
include "##1"}
}}%
\def\psforptips{
\def\PSspeci@l##1##2{{
\d@mx=\psurx bp
\advance \d@mx by -\psllx bp
\divide \d@mx by 1000\multiply\d@mx by \xscale
\incm{\d@mx}
\let\tmpx\dimincm
\d@my=\psury bp
\advance \d@my by -\pslly bp
\divide \d@my by 1000\multiply\d@my by \xscale
\incm{\d@my}
\let\tmpy\dimincm
\d@mx=-\psllx bp
\divide \d@mx by 1000\multiply\d@mx by \xscale
\d@my=-\pslly bp
\divide \d@my by 1000\multiply\d@my by \xscale
\at(\d@mx;\d@my){\special{ps:##1 x=\tmpx, y=\tmpy}}
}}}%
\def\psonlyboxes{
\def\PSspeci@l##1##2{%
\at(0cm;0cm){\boxit{\vbox to\drawinght
  {\vss\hbox to\drawingwd{\at(0cm;0cm){\hbox{({\tt##1})}}\hss}}}}
}}%
\def\psloc@lerr#1{%
\let\savedPSspeci@l=\PSspeci@l%
\def\PSspeci@l##1##2{%
\at(0cm;0cm){\boxit{\vbox to\drawinght
  {\vss\hbox to\drawingwd{\at(0cm;0cm){\hbox{({\tt##1}) #1}}\hss}}}}
\let\PSspeci@l=\savedPSspeci@l
}}%
%
%
\newread\pst@mpin
\newdimen\drawinght\newdimen\drawingwd
\newdimen\psxoffset\newdimen\psyoffset
\newbox\drawingBox
\newcount\xscale \newcount\yscale \newdimen\pscm\pscm=1cm
\newdimen\d@mx \newdimen\d@my
\newdimen\pswdincr \newdimen\pshtincr
\let\ps@nnotation=\relax
{\catcode`\|=0 |catcode`|\=12 |catcode`|
|catcode`#=12 |catcode`*=14
|xdef|backslashother{\}*
|xdef|percentother{
|xdef|tildeother{~}*
|xdef|sharpother{#}*
}%
\def\R@moveMeaningHeader#1:->{}%
\def\uncatcode#1{%
\edef#1{\expandafter\R@moveMeaningHeader\meaning#1}}%
\def\execute#1{#1}
\def\psm@keother#1{\catcode`#112\relax}
\def\executeinspecs#1{%
\execute{\begingroup\let\do\psm@keother\dospecials\catcode`\^^M=9#1\endgroup}}%
\def\@mpty{}%
\def\matchexpin#1#2{
  \fi%
  \edef\tmpb{{#2}}%
  \expandafter\makem@tchtmp\tmpb%
  \edef\tmpa{#1}\edef\tmpb{#2}%
  \expandafter\expandafter\expandafter\m@tchtmp\expandafter\tmpa\tmpb\endm@tch%
  \if\match%
}%
\def\matchin#1#2{%
  \fi%
  \makem@tchtmp{#2}%
  \m@tchtmp#1#2\endm@tch%
  \if\match%
}%
\def\makem@tchtmp#1{\def\m@tchtmp##1#1##2\endm@tch{%
  \def\tmpa{##1}\def\tmpb{##2}\let\m@tchtmp=\relax%
  \ifx\tmpb\@mpty\def\match{YN}%
  \else\def\match{YY}\fi%
}}%
\def\incm#1{{\psxoffset=1cm\d@my=#1
 \d@mx=\d@my
  \divide\d@mx by \psxoffset
  \xdef\dimincm{\number\d@mx.}
  \advance\d@my by -\number\d@mx cm
  \multiply\d@my by 100
 \d@mx=\d@my
  \divide\d@mx by \psxoffset
  \edef\dimincm{\dimincm\number\d@mx}
  \advance\d@my by -\number\d@mx cm
  \multiply\d@my by 100
 \d@mx=\d@my
  \divide\d@mx by \psxoffset
  \xdef\dimincm{\dimincm\number\d@mx}
}}%
%
\newif\ifNotB@undingBox
\newhelp\PShelp{Proceed: you'll have a 5cm square blank box instead of
your graphics (Jean Orloff).}%
\def\s@tsize#1 #2 #3 #4\@ndsize{
  \def\psllx{#1}\def\pslly{#2}%
  \def\psurx{#3}\def\psury{#4}
  \ifx\psurx\@mpty\NotB@undingBoxtrue
  \else
    \drawinght=#4bp\advance\drawinght by-#2bp
    \drawingwd=#3bp\advance\drawingwd by-#1bp
  \fi
  }%
\def\sc@nBBline#1:#2\@ndBBline{\edef\p@rameter{#1}\edef\v@lue{#2}}%
\def\g@bblefirstblank#1#2:{\ifx#1 \else#1\fi#2}%
{\catcode`\%=12
\xdef\B@undingBox{
\def\ReadPSize#1{
 \readfilename#1\relax
 \let\PSfilename=\lastreadfilename
 \openin\pst@mpin=#1\relax
 \ifeof\pst@mpin \errhelp=\PShelp
   \errmessage{I haven't found your postscript file (\PSfilename)}%
   \psloc@lerr{was not found}%
   \s@tsize 0 0 142 142\@ndsize
   \closein\pst@mpin
 \else
   \if\matchexpin{\GlobalInputList}{, \lastreadfilename}%
   \else\xdef\GlobalInputList{\GlobalInputList, \lastreadfilename}%
     \immediate\write\psbj@inaux{\lastreadfilename,}%
   \fi%
   \loop
     \executeinspecs{\catcode`\ =10\global\read\pst@mpin to\n@xtline}%
     \ifeof\pst@mpin
       \errhelp=\PShelp
       \errmessage{(\PSfilename) is not an Encapsulated PostScript File:
           I could not find any \B@undingBox: line.}%
       \edef\v@lue{0 0 142 142:}%
       \psloc@lerr{is not an EPSFile}%
       \NotB@undingBoxfalse
     \else
       \expandafter\sc@nBBline\n@xtline:\@ndBBline
       \ifx\p@rameter\B@undingBox\NotB@undingBoxfalse
         \edef\t@mp{%
           \expandafter\g@bblefirstblank\v@lue\space\space\space}%
         \expandafter\s@tsize\t@mp\@ndsize
       \else\NotB@undingBoxtrue
       \fi
     \fi
   \ifNotB@undingBox\repeat
   \closein\pst@mpin
 \fi
\message{#1}%
}%
%
%
\def\psboxto(#1;#2)#3{\vbox{
   \ReadPSize{#3}%
   \divide\drawingwd by 1000
   \divide\drawinght by 1000
   \d@mx=#1
   \ifdim\d@mx=0pt\xscale=1000
         \else \xscale=\d@mx \divide \xscale by \drawingwd\fi
   \d@my=#2
   \ifdim\d@my=0pt\yscale=1000
         \else \yscale=\d@my \divide \yscale by \drawinght\fi
   \ifnum\yscale=1000
         \else\ifnum\xscale=1000\xscale=\yscale
                    \else\ifnum\yscale<\xscale\xscale=\yscale\fi
              \fi
   \fi
   \divide\pswdincr by 1000 \multiply\pswdincr by \xscale
   \divide\pshtincr by 1000 \multiply\pshtincr by \xscale
   \divide\psxoffset by1000 \multiply\psxoffset by\xscale
   \divide\psyoffset by1000 \multiply\psyoffset by\xscale
   \global\divide\pscm by 1000
   \global\multiply\pscm by\xscale
   \multiply\drawingwd by\xscale \multiply\drawinght by\xscale
   \ifdim\d@mx=0pt\d@mx=\drawingwd\fi
   \ifdim\d@my=0pt\d@my=\drawinght\fi
   \message{scaled \the\xscale}%
 \hbox to\d@mx{\hss\vbox to\d@my{\vss
   \global\setbox\drawingBox=\hbox to 0pt{\kern\psxoffset\vbox to 0pt{
      \kern-\psyoffset
      \PSspeci@l{\PSfilename}{\the\xscale}%
      \vss}\hss\ps@nnotation}%
   \advance\pswdincr by \drawingwd
   \advance\pshtincr by \drawinght
   \global\wd\drawingBox=\the\pswdincr
   \global\ht\drawingBox=\the\pshtincr
   \baselineskip=0pt
   \copy\drawingBox
 \vss}\hss}%
  \global\psxoffset=0pt
  \global\psyoffset=0pt
  \global\pswdincr=0pt
  \global\pshtincr=0pt 
  \global\pscm=1cm 
  \global\drawingwd=\drawingwd
  \global\drawinght=\drawinght
}}%
%
%
\def\psboxscaled#1#2{\vbox{
  \ReadPSize{#2}%
  \xscale=#1
  \message{scaled \the\xscale}%
  \advance\drawingwd by\pswdincr\advance\drawinght by\pshtincr
  \divide\pswdincr by 1000 \multiply\pswdincr by \xscale
  \divide\pshtincr by 1000 \multiply\pshtincr by \xscale
  \divide\psxoffset by1000 \multiply\psxoffset by\xscale
  \divide\psyoffset by1000 \multiply\psyoffset by\xscale
  \divide\drawingwd by1000 \multiply\drawingwd by\xscale
  \divide\drawinght by1000 \multiply\drawinght by\xscale
  \global\divide\pscm by 1000
  \global\multiply\pscm by\xscale
  \global\setbox\drawingBox=\hbox to 0pt{\kern\psxoffset\vbox to 0pt{
     \kern-\psyoffset
     \PSspeci@l{\PSfilename}{\the\xscale}%
     \vss}\hss\ps@nnotation}%
  \advance\pswdincr by \drawingwd
  \advance\pshtincr by \drawinght
  \global\wd\drawingBox=\the\pswdincr
  \global\ht\drawingBox=\the\pshtincr
  \baselineskip=0pt
  \copy\drawingBox
  \global\psxoffset=0pt
  \global\psyoffset=0pt
  \global\pswdincr=0pt
  \global\pshtincr=0pt 
  \global\pscm=1cm
  \global\drawingwd=\drawingwd
  \global\drawinght=\drawinght
}}%
%
\def\psbox#1{\psboxscaled{1000}{#1}}%
\newif\ifn@teof\n@teoftrue
\newif\ifc@ntrolline
\newif\ifmatch
\newread\j@insplitin
\newwrite\j@insplitout
\newwrite\psbj@inaux
\immediate\openout\psbj@inaux=psbjoin.aux
\immediate\write\psbj@inaux{\string\joinfiles}%
\immediate\write\psbj@inaux{\jobname,}%
%
%
\def\toother#1{\ifcat\relax#1\else\expandafter%
  \toother@ux\meaning#1\endtoother@ux\fi}%
\def\toother@ux#1 #2#3\endtoother@ux{\def\tmp{#3}%
  \ifx\tmp\@mpty\def\tmp{#2}\let\next=\relax%
  \else\def\next{\toother@ux#2#3\endtoother@ux}\fi%
\next}%
%
%
\let\readfilenamehook=\relax
\def\re@d{\expandafter\re@daux}
\def\re@daux{\futurelet\nextchar\stopre@dtest}%
\def\re@dnext{\xdef\lastreadfilename{\lastreadfilename\nextchar}%
  \afterassignment\re@d\let\nextchar}%
\def\stopre@d{\egroup\readfilenamehook}%
\def\stopre@dtest{%
  \ifcat\nextchar\relax\let\nextread\stopre@d
  \else
    \ifcat\nextchar\space\def\nextread{%
      \afterassignment\stopre@d\chardef\nextchar=`}%
    \else\let\nextread=\re@dnext
      \toother\nextchar
      \edef\nextchar{\tmp}%
    \fi
  \fi\nextread}%
\def\readfilename{\vbox\bgroup%
  \let\\=\backslashother \let\%=\percentother \let\~=\tildeother
  \let\#=\sharpother \xdef\lastreadfilename{}%
  \re@d}%
%
%
\xdef\GlobalInputList{\jobname}%
\def\psnewinput{%
  \def\readfilenamehook{
    \if\matchexpin{\GlobalInputList}{, \lastreadfilename}%
    \else\xdef\GlobalInputList{\GlobalInputList, \lastreadfilename}%
      \immediate\write\psbj@inaux{\lastreadfilename,}%
    \fi%
    \ps@ldinput\lastreadfilename\relax%
    \let\readfilenamehook=\relax%
  }\readfilename%
}%
\expandafter\ifx\csname @@input\endcsname\relax    
  \immediate\let\ps@ldinput=\input\def\input{\psnewinput}%
\else
  \immediate\let\ps@ldinput=\@@input
  \def\@@input{\psnewinput}%
\fi%
\def\nowarnopenout{%
 \def\warnopenout##1##2{%
   \readfilename##2\relax
   \message{\lastreadfilename}%
   \immediate\openout##1=\lastreadfilename\relax}}%
\def\warnopenout#1#2{%
 \readfilename#2\relax
 \def\t@mp{TrashMe,psbjoin.aux,psbjoint.tex,}\uncatcode\t@mp
 \if\matchexpin{\t@mp}{\lastreadfilename,}%
 \else
   \immediate\openin\pst@mpin=\lastreadfilename\relax
   \ifeof\pst@mpin
     \else
     \errhelp{If the content of this file is so precious to you, abort (ie
press x or e) and rename it before retrying.}%
     \errmessage{I'm just about to replace your file named \lastreadfilename}%
   \fi
   \immediate\closein\pst@mpin
 \fi
 \message{\lastreadfilename}%
 \immediate\openout#1=\lastreadfilename\relax}%
{\catcode`\%=12\catcode`\*=14
\gdef\splitfile#1{*
 \readfilename#1\relax
 \immediate\openin\j@insplitin=\lastreadfilename\relax
 \ifeof\j@insplitin
   \message{! I couldn't find and split \lastreadfilename!}*
 \else
   \immediate\openout\j@insplitout=TrashMe
   \message{< Splitting \lastreadfilename\space into}*
   \loop
     \ifeof\j@insplitin
       \immediate\closein\j@insplitin\n@teoffalse
     \else
       \n@teoftrue
       \executeinspecs{\global\read\j@insplitin to\spl@tinline\expandafter
         \ch@ckbeginnewfile\spl@tinline
       \ifc@ntrolline
       \else
         \toks0=\expandafter{\spl@tinline}*
         \immediate\write\j@insplitout{\the\toks0}*
       \fi
     \fi
   \ifn@teof\repeat
   \immediate\closeout\j@insplitout
 \fi\message{>}*
}*
\gdef\ch@ckbeginnewfile#1
 \def\t@mp{#1}*
 \ifx\@mpty\t@mp
   \def\t@mp{#3}*
   \ifx\@mpty\t@mp
     \global\c@ntrollinefalse
   \else
     \immediate\closeout\j@insplitout
     \warnopenout\j@insplitout{#2}*
     \global\c@ntrollinetrue
   \fi
 \else
   \global\c@ntrollinefalse
 \fi}*
\gdef\joinfiles#1\into#2{*
 \message{< Joining following files into}*
 \warnopenout\j@insplitout{#2}*
 \message{:}*
 {*
 \edef\w@##1{\immediate\write\j@insplitout{##1}}*
\w@{
\w@{
\w@{
\w@{
\w@{
\w@{
\w@{
\w@{
\w@{
\w@{
\w@{\string\input\space psbox.tex}*
\w@{\string\splitfile{\string\jobname}}*
\w@{\string\let\string\autojoin=\string\relax}*
}*
 \expandafter\tre@tfilelist#1, \endtre@t
 \immediate\closeout\j@insplitout
 \message{>}*
}*
\gdef\tre@tfilelist#1, #2\endtre@t{*
 \readfilename#1\relax
 \ifx\@mpty\lastreadfilename
 \else
   \immediate\openin\j@insplitin=\lastreadfilename\relax
   \ifeof\j@insplitin
     \errmessage{I couldn't find file \lastreadfilename}*
   \else
     \message{\lastreadfilename}*
     \immediate\write\j@insplitout{
     \executeinspecs{\global\read\j@insplitin to\oldj@ininline}*
     \loop
       \ifeof\j@insplitin\immediate\closein\j@insplitin\n@teoffalse
       \else\n@teoftrue
         \executeinspecs{\global\read\j@insplitin to\j@ininline}*
         \toks0=\expandafter{\oldj@ininline}*
         \let\oldj@ininline=\j@ininline
         \immediate\write\j@insplitout{\the\toks0}*
       \fi
     \ifn@teof
     \repeat
   \immediate\closein\j@insplitin
   \fi
   \tre@tfilelist#2, \endtre@t
 \fi}*
}%
\def\autojoin{%
 \immediate\write\psbj@inaux{\string\into{psbjoint.tex}}%
 \immediate\closeout\psbj@inaux
 \expandafter\joinfiles\GlobalInputList\into{psbjoint.tex}%
}%
%
%
%
\def\centinsert#1{\midinsert\line{\hss#1\hss}\endinsert}%
\def\psannotate#1#2{\vbox{%
  \def\ps@nnotation{#2\global\let\ps@nnotation=\relax}#1}}%
\def\pscaption#1#2{\vbox{%
   \setbox\drawingBox=#1
   \copy\drawingBox
   \vskip\baselineskip
   \vbox{\hsize=\wd\drawingBox\setbox0=\hbox{#2}%
     \ifdim\wd0>\hsize
       \noindent\unhbox0\tolerance=5000
    \else\centerline{\box0}%
    \fi
}}}%
%
\def\at(#1;#2)#3{\setbox0=\hbox{#3}\ht0=0pt\dp0=0pt
  \rlap{\kern#1\vbox to0pt{\kern-#2\box0\vss}}}%
%
\newdimen\gridht \newdimen\gridwd
\def\gridfill(#1;#2){%
  \setbox0=\hbox to 1\pscm
  {\vrule height1\pscm width.4pt\leaders\hrule\hfill}%
  \gridht=#1
  \divide\gridht by \ht0
  \multiply\gridht by \ht0
  \gridwd=#2
  \divide\gridwd by \wd0
  \multiply\gridwd by \wd0
  \advance \gridwd by \wd0
  \vbox to \gridht{\leaders\hbox to\gridwd{\leaders\box0\hfill}\vfill}}%
%
\def\fillinggrid{\at(0cm;0cm){\vbox{%
  \gridfill(\drawinght;\drawingwd)}}}%
%
%
\def\textleftof#1:{%
  \setbox1=#1
  \setbox0=\vbox\bgroup
    \advance\hsize by -\wd1 \advance\hsize by -2em}%
\def\textrightof#1:{%
  \setbox0=#1
  \setbox1=\vbox\bgroup
    \advance\hsize by -\wd0 \advance\hsize by -2em}%
\def\endtext{%
  \egroup
  \hbox to \hsize{\valign{\vfil##\vfil\cr%
\box0\cr%
\noalign{\hss}\box1\cr}}}%
%
\def\frameit#1#2#3{\hbox{\vrule width#1\vbox{%
  \hrule height#1\vskip#2\hbox{\hskip#2\vbox{#3}\hskip#2}%
        \vskip#2\hrule height#1}\vrule width#1}}%
\def\boxit#1{\frameit{0.4pt}{0pt}{#1}}%
\catcode`\@=12 
%
 \psfordvips   